# Dual-mode operation of an optical lattice clock using strontium and ytterbium atoms

Daisuke Akamatsu, Takumi Kobayashi, Yusuke Hisai, Takehiko Tanabe, Kazumoto Hosaka, Masami Yasuda, and Feng-Lei Hong

*Abstract*— We have developed an optical lattice clock that can operate in dual modes: a strontium (Sr) clock mode and an ytterbium (Yb) clock mode. Dual-mode operation of the Sr-Yb optical lattice clock is achieved by alternately cooling and trapping $^{87}$Sr and $^{171}$Yb atoms inside the vacuum chamber of the clock. Optical lattices for Sr and Yb atoms were arranged with horizontal and vertical configurations, respectively, resulting in a small distance of the order of 100 μm between the trapped Sr and Yb atoms. The $^1S_0$-$^3P_0$ clock transitions in the trapped atoms were interrogated in turn and the clock lasers were stabilized to the transitions. We demonstrated the frequency ratio measurement of the Sr and Yb clock transitions by using the dual-mode operation of the Sr-Yb optical lattice clock. The dual-mode operation can reduce the uncertainty of the blackbody radiation shift in the frequency ratio measurement, because both Sr and Yb atoms share the same blackbody radiation.

## I. INTRODUCTION

THE second as a unit of time is defined based on the microwave transition in Cs and can be realized with $10^{-16}$-level accuracy by a Cs fountain clock. Optical clocks, such as optical lattice clocks and single ion clocks, have surpassed microwave clocks and are expected to realize a new definition of the second [1, 2]. Optical lattice clocks (OLC), which employ optical transitions of neutral atoms in an optical lattice at a magic wavelength as a frequency reference, were proposed by H. Katori in 2001 [3]. A Sr OLC [4] was demonstrated in 2005, and an Yb OLC [5,6] and a Hg OLC [7] were also subsequently demonstrated in 2009 and in 2012, respectively. Of these OLCs, Sr and Yb OLCs have been the most extensively studied. Absolute frequency measurements of the clock transition in Sr have been reported by 7 institutes [8-14] and that in Yb by 4 institutes [6,15-17]. The measurement results are mainly limited by the accuracy of the microwave frequency standards used in the experiments. A frequency ratio measurement by the direct frequency comparison of two independent optical clocks is free from this limitation. The frequency ratio of Sr and Yb clock transitions has been measured by groups at NMIJ [18], RIKEN [19,20] and PTB/INRiM [21]. The measured absolute frequencies and frequency ratios have been reported to the International Committee for Weights and Measures (CIPM) via the Consultative Committee for Time and Frequency (CCTF) and contributed to discussions aimed at the redefinition of the second.

The RIKEN group demonstrated a compatible optical lattice clock. They converted an optical lattice clock from Sr to Yb by changing the lights coupled to the fibers, which transmitted the lights for clock operation to a chamber [19].

Here we report the demonstration of the dual-mode operation of an optical lattice clock of Sr and Yb by overlapping all the cooling lasers with dichromatic mirrors. Thanks to the similar vapor pressures of Sr and Yb, we can alternately laser-cool and trap the atomic species in the same chamber without changing the temperature of an atomic oven containing a mixture of Sr and Yb. We alternately interrogate each clock transition and stabilize both clock lasers to the clock transitions. We measure the frequency ratio of the clock transition frequencies, which agrees well with previous measurements. Since the atoms are affected by the same blackbody radiation during the ratio measurement experiment, the AC Stark shift due to the blackbody radiation can be partially cancelled out in the ratio measurement, resulting in a smaller uncertainty in the error budget.

## II. EXPERIMENTAL APPARATUS

### A. Cooling and trapping lasers for strontium

Lasers for the 1$^{st}$ and 2$^{nd}$ stage magneto-optical trapping (MOT) of Sr and Yb are shown schematically in Fig. 1.

A laser beam at 461 nm was generated by the second harmonic generation of 922 nm light from an external cavity diode laser (ECDL) [22]. The light from the ECDL was amplified to about 1 W by a tapered amplifier (TA#1 in Fig. 1). The amplified light was mainly coupled to a periodically poled lithium niobate (PPLN) waveguide (PPLNwg#1) and frequency-doubled to generate light at 461 nm. Part of the 461-nm light from PPLNwg#1 was employed for the spectroscopy

D. Akamatsu, T. Kobayashi, T. Tanabe, K. Hosaka, and M. Yasuda are with the Time Standards Group, National Metrology Institute of Japan, National Institute of Advanced Industrial Science and Technology, Tsukuba 305-8563, Japan (e-mail: d-akamatsu@aist.go.jp).

Y. Hisai and F. -L. Hong are with the Department of Physics, Graduate School of Engineering Science, Yokohama National University, Yokohama 240-8501, Japan.
This work was supported by JSPS KAKENHI Grant Numbers JP24740281, JP15K05238, JP17H01151.

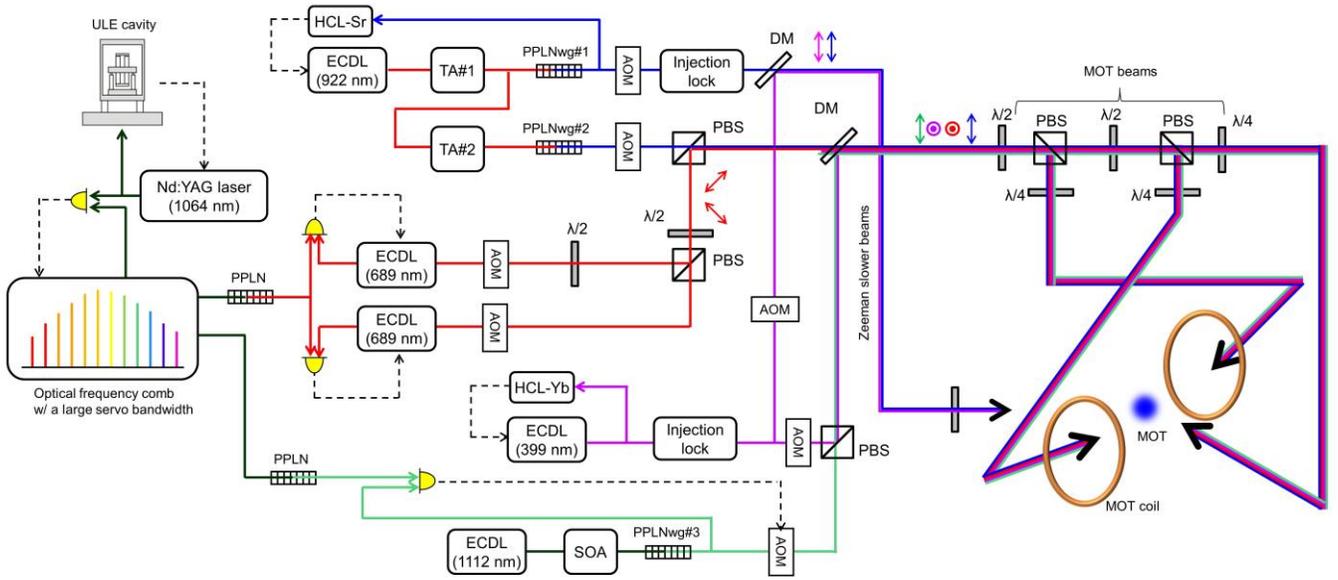

Fig. 1. Schematic of experimental setup. The bronze rings represent the anti-Helmholtz coils (MOT coils) used for magneto-optical trapping. The lattice geometries are different for Sr and Yb: a vertical optical lattice is employed for Sr and a nearly horizontal optical lattice for Yb. ECDL, extended cavity diode laser; TA, tapered amplifier; PPLNwg, periodically poled lithium niobate waveguide; SOA, semiconductor optical amplifier; ULE cavity, ultra-low-expansion cavity; AOM, acousto-optic modulator.

of Sr using a hollow cathode lamp (HCL-Sr in Fig. 1). The spectroscopic signal was used to stabilize the ECDL frequency at 922 nm. The rest of the beam was frequency-shifted by an acousto-optic modulator (AOM), so that its frequency was red-detuned by 366 MHz from the $^1S_0$-$^1P_1$ transition in Sr (1$^{st}$ stage cooling transition). The beam was finally used to injection-lock a blue laser diode (Nichia NDB4216E), which provided 50 mW of power for Zeeman slowing.

Part of the TA#1 output was injected into TA#2. The TA#2 output was coupled to another PPLN waveguide (PPLNwg#2), and the generated 461-nm light was used for the 1$^{st}$ stage MOT of Sr (Blue MOT). The Blue MOT beam was red-detuned by 43 MHz by using another AOM. To close the 1$^{st}$ stage cooling cycle, two repumping lasers at 679 and 707 nm were also used [23] (not shown in Fig.1).

The lights at 689 nm for the 2$^{nd}$ stage MOT of Sr (Red MOT) were generated by ECDLs. One of the lights (a trapping laser) drove the $^1S_0$ ($F_g$=9/2) - $^3P_1$ ($F_e$=11/2) transition and created a position-dependent force and the other (a stirring laser) drove the $F_g$=9/2 - $F_e$=9/2 transition and mixed the magnetic sublevels of the ground states [24]. Since the linewidth of the $^3P_1$ state is as narrow as 7.6 kHz, the linewidth of both lasers must be narrow. We employed a linewidth transfer method to reduce the laser linewidth, the details of which were published in [25]. The trapping and stirring lasers were overlapped with a polarizing beam splitter (PBS). Then the polarizations were rotated 45 degrees by using a half-waveplate. Again, the 689-nm lights were overlapped with the light at 461 nm by using a broadband PBS.

The lattice laser at 813 nm was generated by a Ti:sapphire laser (TekhnoScan TIS-SF-07). To reduce the short-term frequency noise or frequency jitter of the lattice laser, we employed a Fabry-Perot cavity as a reference cavity that had a finesse of about 200. The laser frequency was stabilized at the shoulder of the cavity transmission by using a tweeter mirror and the thick etalon of the Ti:sapphire laser. To reduce the long-term frequency drift and fix the absolute frequency, we stabilized the laser to our optical frequency comb referenced to the coordinated universal time at NMIJ (UTC(NMIJ)) by controlling the Fabry-Perot cavity. The optical lattice for Sr was set in a vertical configuration operating at $f_{\text{Sr-lattice}}$=368.554 42 (15) THz. The lattice trap depth was 24 $E_r$, where $E_r$ is the photon recoil energy. The lattice photon recoil energy for Sr is $E_r/k_B$=165 nK.

### B. Cooling and trapping lasers of ytterbium

The light at 399 nm was generated by a home-built ECDL [26] and the frequency was stabilized using an Yb hollow cathode lamp. The light was used to injection-lock a high-power laser diode (Nichia NDV4B16) [27]. The light was split into two beams, one of which was used for Zeeman slowing and the other for the 1$^{st}$ stage MOT of Yb (Violet MOT). The detunings of the lights for the Violet MOT and Zeeman slowing were -15 and -278 MHz, respectively.

The 556-nm light was produced by frequency-doubling the light from a commercial cateye ECDL at 1112 nm (MOGlab CEL002). The output of the ECDL was amplified to 200 mW by a semiconductor optical amplifier. The amplified light was frequency-doubled by using a PPLN waveguide (PPLNwg#3). Part of the green light was used to detect the beat signal with the frequency comb for linewidth transfer. The light was phase-locked to the frequency comb using an AOM. The slow frequency drift was compensated for by using the PZT of the cateye ECDL. A broadband PBS was again used to overlap the light for the 2$^{nd}$ stage MOT of Yb (Green MOT) with that for the Violet MOT.

A lattice laser emitting at 759 nm was obtained by using another commercially available Ti:sapphire laser (Msquare SolsTiS-SRX). The linewidth of the lattice laser was less than

50 kHz and its frequency was stabilized to the frequency comb, which was also used for stabilizing the lattice laser for Sr. The optical lattice for Yb was set in a nearly horizontal configuration operating at $f_{Yb\text{-lattice}}$=394.798 30 (15) THz. The lattice trap depth was $877E_r$, where the lattice photon recoil energy was $E_r/k_B$=97 nK.

*C. "Master" narrow linewidth laser and clock lasers for strontium and ytterbium*

A Nd:YAG laser operating at 1064 nm, which was stabilized to a high finesse cavity, was employed as a "master" laser for narrow linewidth lasers, such as the 2nd stage MOT lasers and clock lasers for Sr and Yb. The Allan deviation of the master laser was estimated to be about $2 \times 10^{-15}$ from 1 to 50 s. The clock lasers for Sr and Yb were also prepared by the linewidth transfer method with a high-speed controllable optical frequency comb. It should be noted that since both clock lasers were phase-locked to the same frequency comb, the short-term noise would be partially cancelled out during the simultaneous operation of the optical lattice clocks as demonstrated in [20].

The clock lasers for Sr and Yb were delivered to a vacuum chamber after passing through AOMs to control the clock laser frequencies. Since the RF power driving the AOMs was small enough, we neglected the systematic uncertainties due to the phase drift effects. The atoms in the lattice were illuminated by the clock lasers through curved mirrors, which were used to form 1D optical lattices.

### III. DUAL-MODE OPERATION OF SR-YB OPTICAL LATTICE CLOCK

The laser beams for the MOT and Zeeman slowing of Sr and Yb were overlapped with dedicated dichroic mirrors. The polarizations of the overlapped MOT lights were rotated by 45 degrees by a broadband half-waveplate, and then a broadband PBS split the beam into two beams. One of the beams was used as a horizontal cooling light, and the other was again split into two with another pair consisting of a broadband half-waveplate and a PBS. The beams travelled diagonally upwards in the vacuum chamber. These three beams were all retro-reflected and formed three pairs of counter-propagating beams for MOT (not shown in Fig. 1).

The temperature of the atomic oven containing the Sr and Yb was set at about 470 °C, where the vapor pressure of the Sr and Yb were 0.3 and 1 Pa, respectively. To collimate the atomic beam, a honeycomb nozzle was attached to the exit of the atomic oven. The magnetic field for Zeeman slowing was kept on during the experiment. Commercially available current sources were used to generate the quadrupole magnetic field for MOT and were switched by an electric circuit with insulated gate bipolar transistors. The quadrupole field gradient for the Blue and Violet MOT was 35 mT/m, that for the Red MOT was 3 mT/m, that for the Green MOT was 11 mT/m.

The timing chart of the dual-mode clock operation is shown in Fig. 2. By alternately switching on the lasers and the quadrupole field, the atomic species were laser-cooled, trapped and probed by turns.

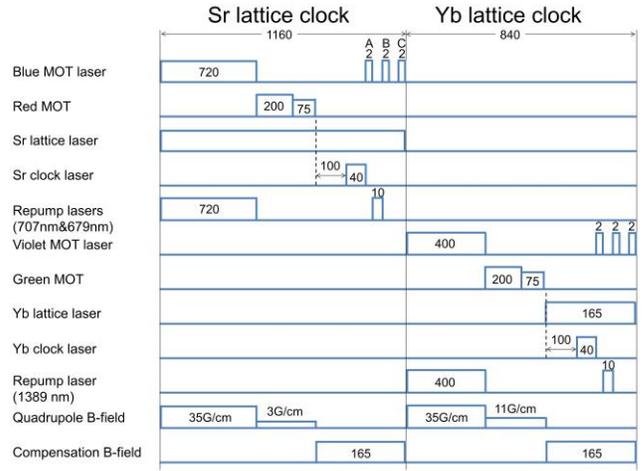

Fig. 2. Timing chart for dual-mode operation of optical lattice clocks (time unit: ms).

After 720 ms of the Blue MOT stage, the 689-nm lasers (both trapping and stirring lasers) were turned on. For the first 200 ms of the Red MOT, the trapping (stirring) laser frequency was modulated at 30 kHz with a spectral bandwidth of 2.3 (3.4) MHz and detuned by -1.2 (-1.8) MHz to increase the capture velocity range. After the broadband stage, single mode frequency cooling was applied by lowering the power, terminating the modulation, and decreasing the detunings to -200 and -100 kHz for the trapping and stirring lasers, respectively. Finally, $10^5$ atoms were cooled to a few µK. The lattice laser was kept on during the MOT stages. We waited 100 ms before introducing the clock laser in order to let the untrapped atoms fall and the magnetic field caused by the eddy current cease. The compensation coil was turned on in order to zero the magnetic field at the atomic cloud during the interrogation of the clock transition. With the zero magnetic field, all the Zeeman sublevels of the clock transition were degenerate. We obtained the fluorescence images (A, B, C) by turning on the Blue MOT laser for 2 ms. Image A provided the number of atoms in the $^1S_0$ state. Image B, which was obtained after the repumping pulse, provided the number of atoms in the $^3P_0$ state. Image C was used to subtract the background from the images of A and B.

After the Sr lattice clock stage, the violet laser was turned on to collect Yb atoms. After 400 ms of the Violet MOT stage, the 556-nm laser was turned on. As with Sr, during the first 200 ms of the Green MOT stage, the laser spectrum was broadened to 5.5 MHz by frequency modulation. Then during the last 75 ms, the frequency modulation was terminated and the detuning was reduced to 2.0 MHz for deep cooling. Unlike Sr, the irradiation of the lattice laser during the Green MOT stage resulted in reduction of the number of the atoms in the lattice. We therefore turned on the lattice laser after the Green MOT laser was off. After 100 ms, we irradiated the atoms with the clock laser. The three images, A, B, and C were obtained again and employed to determine the excitation probabilities.

Figure 3 shows an overlapped image of the fluorescence images A of Sr and Yb obtained with an electron multiplying charge-coupled device (EMCCD) camera. The EMCCD camera was situated at approximately the same height as the horizontal Yb optical lattice, and the angle between the lattice beam and the sight line of the camera was 30 degrees. The vertically-long cloud is the Sr atoms trapped in the vertical lattice and horizontally-long one is the Yb cloud in the horizontal lattice. The clouds are separated by a distance of several 100 µm. The numbers of the trapped atoms of Sr and Yb are estimated to be ~$7 \times 10^3$ and $2 \times 10^3$, respectively. Figure 4(a) shows a typical spectrum of the degenerate clock transition of Sr for an interrogation time of 40 ms. The observed linewidth of 35 Hz was almost the Fourier limited linewidth. The noisier spectrum of Yb (Fig. 4(b)) is due to there being a smaller number of trapped atoms in the lattice.

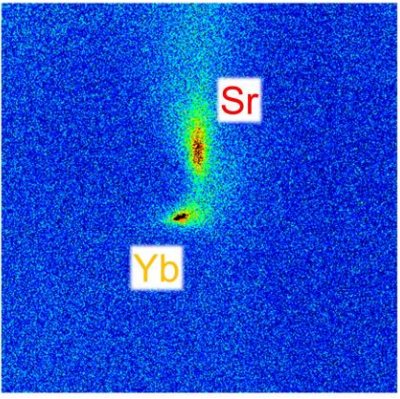

Fig. 3. Fluorescence image of Sr and Yb. The images of Sr and Yb were obtained separately and superimposed digitally.

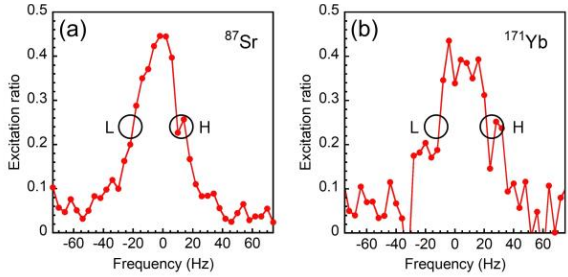

Fig. 4. Spectrum of the clock transition of (a) the Sr OLC and (b) the Yb OLC. H and L are the high- and low-frequency points at the full-width at half maximum of each spectrum.

For the clock operation, the clock lasers were frequency-stabilized to both degenerate spectra using the AOMs. Each clock laser alternately probed the high- and low-frequency points (H and L in Fig. 4) at the full width of half maximum of each spectrum. The cycle time for probing the clock transitions of Sr and Yb was 2 seconds. The RF frequency driving AOMs were controlled to equalize the excitation probabilities. The average frequency of the high- and low-frequency points determined the center frequency of the clock transition.

To demonstrate the dual-mode clock operation, we performed a frequency ratio measurement of Sr and Yb using the developed optical lattice clock. Figures 5 (a) and (b) show a typical Allan standard deviation of the measured frequency ratio and four frequency ratio measurement results. Thanks to the improvement made on the short-term stability of the master laser, the short-term stability of each measurement was superior to our previous measurements [16]. The Allan deviation reached $6 \times 10^{-16}$ after 3000 s of averaging time. The error bar for the ratio measurement result was determined by the Allan standard deviation at the longest averaging time for each measurement. The fractional statistical uncertainty of the ratio measurement was calculated to be $8.3 \times 10^{-16}$.

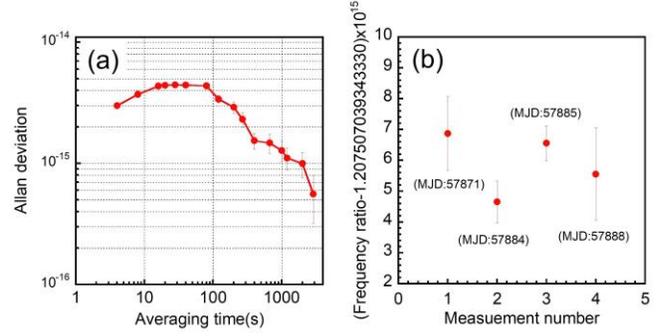

Fig. 5. (a) Allan standard deviation of the frequency ratio measurement for measurement number 3 in Fig. 5 (b) Frequency ratio measurements of Sr and Yb optical lattice clocks. Data shown in this figure do not include the systematic corrections.

Here, we conservatively evaluated the corrections and their uncertainties from previous reports. Since we employed unpolarized atoms, the asymmetric distribution of the population and residual magnetic field (0.7 µT in this experiment) can cause a 1st order Zeeman shift of the clock transition [28]. We calculated the uncertainties due to this effect to be $4.6 \times 10^{-15}$ and $1.6 \times 10^{-15}$ from the 1st order Zeeman coefficients for Sr [29] and Yb [6], respectively. The correction of the AC Stark shift of the Yb clock transition caused by the lattice laser was calculated to be $0.3(3.2) \times 10^{-15}$ based on [30]. From previous measurements of the temperature coefficients [31-33], the correction due to the ac Stark shift of the clock transition of Sr and Yb induced by blackbody radiation were calculated to be $5.5(0.2) \times 10^{-15}$ and $2.6(0.1) \times 10^{-15}$ for an environmental temperature of $T$=303(3) K. The discussion section provides a detailed evaluation of the blackbody radiation shift with a smaller uncertainty taking account of the fact that Sr and Yb atoms share the same environmental temperature. Other corrections and uncertainties are below $1 \times 10^{-15}$.

With the above corrections, we determined the frequency ratio to be 1.207 507 039 343 332 4 (1 0)$_{stat}$ (7 1)$_{sys}$, with a total fractional uncertainty of $5.9 \times 10^{-15}$. The measurement result agrees with a weighted average of previous experimental results [18-21] within our measurement uncertainty.

## IV. DISCUSSIONS AND CONCLUSIONS

In this experiment, we employed the degenerate Zeeman spectrum of the unpolarized atoms, which caused the relatively large uncertainty due to the 1st order Zeeman shift. The uncertainty of the frequency ratio measurement can be improved by employing spin-polarized atoms and cancelling the 1st order Zeeman shift [34]. The state of the art results as low as $5\times 10^{-18}$ have been published [35], demonstrating the capacity of optical lattice clocks to control the systematics in the low $10^{-18}$ range. The in-situ monitoring of the blackbody radiation environment is still challenging and the temperature measurement results in a leading uncertainty in most experiments [10-13].

The frequency ratio measured in an environment with a temperature $T$ is

$$r(T) = \frac{\nu^{Yb}(T=0)+\Delta\nu^{Yb}}{\nu^{Sr}(T=0)+\Delta\nu^{Sr}} \quad (1),$$

$$= \frac{\nu^{Yb}(T=0)}{\nu^{Sr}(T=0)}\left(1+\frac{\Delta\nu^{Yb}}{\nu^{Yb}(T=0)}-\frac{\Delta\nu^{Sr}}{\nu^{Sr}(T=0)}+O\left(\left(\frac{\Delta\nu}{\nu}\right)^2\right)\right)$$

where $\Delta\nu^{Sr(Yb)}$ is the AC Stark shift of Sr (Yb) clock transition caused by the blackbody radiation. The second and the subsequent terms in parentheses on the second line are the shift in the ratio measurement at $T$. In the following, we neglect the last term in parentheses. The fractional blackbody radiation shift of Sr(Yb) can be written as

$$\frac{\Delta\nu^{Sr(Yb)}}{\nu^{Sr(Yb)}(T=0)} = \frac{\Delta\nu^{Sr(Yb)}_{stat}}{\nu^{Sr(Yb)}(T=0)}\left(\frac{T^{Sr(Yb)}}{300\text{ (K)}}\right)^4 + \frac{\Delta\nu^{Sr(Yb)}_{dyn}}{\nu^{Sr(Yb)}(T=0)}\left(\frac{T^{Sr(Yb)}}{300\text{ (K)}}\right)^6 \quad (2),$$

where $\Delta\nu^{Sr(Yb)}_{stat}$ and $\Delta\nu^{Sr(Yb)}_{dyn}$ are the static and dynamic shifts of Sr (Yb), respectively. $T^{Sr(Yb)}$ is the environmental temperature of Sr (Yb) optical lattice clock. We neglected the higher order of the dynamical shift. Since both atoms were in the temperature environment $T^{Dual}$ in this experiment, the fractional shift in the frequency ratio measurement can be written as

$$\frac{\Delta\nu^{Yb}}{\nu^{Yb}(T=0)}-\frac{\Delta\nu^{Sr}}{\nu^{Sr}(T=0)} = \left(\frac{\Delta\nu^{Yb}_{stat}}{\nu^{Yb}(T=0)}-\frac{\Delta\nu^{Sr}_{stat}}{\nu^{Sr}(T=0)}\right)\left(\frac{T^{Dual}}{300\text{ (K)}}\right)^4 \quad (3).$$
$$+ \left(\frac{\Delta\nu^{Yb}_{dyn}}{\nu^{Yb}(T=0)}-\frac{\Delta\nu^{Sr}_{dyn}}{\nu^{Sr}(T=0)}\right)\left(\frac{T^{Dual}}{300\text{ (K)}}\right)^6$$

We calculated the fractional correction due to the blackbody radiation to be $29.7(1.2)\times 10^{-16}$ using the common environmental temperature of $T^{Dual}=303(3)$ K. It should be noted that the uncertainties on the coefficients are negligible here. The obtained uncertainty is about one half of that calculated based on (1) for the non-common environment temperatures $T^{Sr}=303(3)$ K and $T^{Yb}=303(3)$ K. In our system, owing to the heating of the MOT and Zeeman slower coils, the chamber has a relatively large temperature gradient. It is possible to reduce the uncertainty of the temperature to 1 K by cooling the coil more efficiently and carefully monitoring the chamber temperature. The measurement uncertainty can reach $3.9\times 10^{-17}$ for $T^{Dual}=297(1)$ K.

The simultaneous operation of the Sr and Yb optical lattice clocks is very attractive since it would improve the short-term stability of the measurement by cancelling the clock laser noise [20], and result in a shorter averaging time to reach the same statistical uncertainty. Furthermore, besides the BBR cancellation, the uncertainty of the second order Zeeman shift correction would also be partially cancelled since the atoms share the same magnetic field.

For simultaneous operation, the effects of the cooling and trapping lights used for Sr (Yb) on Yb (Sr) must be considered. In the 1st MOT stage, we observed a decrease in the number of trapped Sr atoms in the simultaneous MOT of Sr and Yb. This is because Sr in the $^1P_1$ state was ionized by the 399-nm Yb MOT laser. The observed atomic loss was only a few percent, which is not a very serious concern as regards simultaneous operation. In the 2nd MOT stage, although the magnetic field gradients of the Red MOT stage (3 mT/m) and the Green MOT stage (11 mT/m) were different in this experiment, we found it was possible to implement Green MOT with a magnetic field gradient of 3 mT/m without changing the number of atoms.

Another concern is the ac Stark shift caused by the lattice beam for another species. In our system, the two lattice beams are not overlapped to avoid the beams illuminating other species. However, the wing of the lattice beams and the scattered stray light in the chamber may cause an AC Stark shift. In such a case, an interleaved operation, where the lattice beam for the other turns on and off, would enable the effect to be evaluated.

In this paper, we have demonstrated the first dual mode operation of a Sr-Yb optical lattice clock. We successfully measured the frequency ratio of Yb/Sr with the optical lattice clock. The measurement result agrees with those of previous reports. The relatively larger uncertainty in the present measurement is due to the first order Zeeman shift and the uncertainty in the magic wavelength of Yb. These effects can be reduced by interrogating the spin-polarized atoms under an intentionally applied magnetic field and evaluating the uncertainty effects on site. The dual mode operation can reduce the uncertainty of the blackbody radiation shift in the frequency ratio measurement, because Sr and Yb atoms share the same blackbody radiation, which allows an uncertainty of $4\times 10^{-17}$ for a temperature measurement uncertainty of 1 K at room temperature.

ACKNOWLEDGMENT

We thank Hajime Inaba and Sho Okubo for technical assistance with the fiber frequency combs and Masatoshi Kajita for valuable comments on the cancellation of the Zeeman shift. We appreciate the loan of the 1112 nm laser system from Takuya Kohno of Gifu College.